# 5-axis High Speed Milling Optimisation

**Christophe Tournier [1], Sylvain Lavernhe [1], Claire Lartigue [1,2]**

[1] *Laboratoire Universitaire de Recherche en Production Automatisée*
 *ENS de Cachan - Université Paris Sud 11*
*61 Avenue du Président Wilson, 94235 Cachan Cedex - France*

[2] *IUT de Cachan - Université Paris Sud 11*
*9 Avenue de la division Leclerc, 94234 Cachan Cedex – France*

*Email: name@lurpa.ens-cachan.fr*

ABSTRACT: *Manufacturing of free form parts relies on the calculation of a tool path based on a CAD model, on a machining strategy and on a given numerically controlled machine tool. In order to reach the best possible performances, it is necessary to take into account a maximum of constraints during tool path calculation. For this purpose, we have developed a surface representation of the tool paths to manage 5-axis High Speed Milling, which is the most complicated case. This model allows integrating early in the step of tool path computation the machine tool geometrical constraints (axis ranges, part holder orientation), kinematical constraints (speed and acceleration on the axes, singularities) as well as gouging issues between the tool and the part. The aim of the paper is to optimize the step of 5-axis HSM tool path calculation with a bi-parameter surface representation of the tool path. We propose an example of integration of the digital process for tool path computation, ensuring the required quality and maximum productivity.*

RÉSUMÉ : *la réalisation des pièces de formes complexes passe par la génération de trajectoires d'usinage, basée sur un modèle CAO, une stratégie d'usinage et une machine outil à commande numérique donnée. Afin d'assurer les meilleures performances possibles en terme de qualité et de productivité, il est nécessaire d'intégrer un maximum de contraintes lors de la génération des trajets d'usinage. Pour répondre au cas complexe du fraisage 5 axes grande vitesse, nous avons développé un modèle surfacique de représentation des trajectoires d'usinage. Ce modèle permet d'intégrer lors du calcul des trajectoires les contraintes géométriques de la machine (dépassement des courses, position du porte-pièce), cinématiques (vitesses et accélérations des axes, singularités), ainsi que les problèmes de collisions avec la partie active de l'outil. Ainsi cet article est consacré à l'optimisation de l'usinage 5 axes des pièces de formes complexes en grande vitesse à l'aide du modèle surfacique de représentation des trajectories. Nous présentons un exemple d'intégration de la chaîne numérique dans la recherche de trajectoires d'usinage assurant une qualité et une productivité maximum.*

MOTS-CLÉS : *Chaîne numérique, fraisage 5 axes, surfaces complexes, UGV, post-processeur.*

## 1. Introduction

Free form manufacturing process by machining is crucial within the fields of aeronautics, automotive, molds and dies industries. Free form machining relies on a step of tool path computation from a CAD model, a machining strategy and a numerically controlled machine tool. This process largely evolved these last years by the use of new techniques such as the digital mock-up (i.e. the complete modeling of the product and its process in a CAD/CAM system), by the use of new interpolation formats within the numerical controller (NC) of the machine tool and by High Speed Machining (HSM) [Duc 1998].

As each stage of the process relies on its own modeling of the surface geometry, approximations occur during CAD/CAM activities and during trajectory follow up by the NC unit.

The step of tool path computation in the CAM system is based on a CAD model, a tool geometry, a machining strategy (a machining direction and sampling parameters) and a tool path description format. Whatever the type of machining, 3 or 5-axes, and the description format, the calculated tool path transmitted to the NC unit is a set of characteristic points: tool center locations in the case of linear interpolation and control points in the case of polynomial interpolation.

The use of HSM makes it possible to limit polishing operations. However, NC unit behaviors and dynamic phenomena during machining happen to be modified: difficulty in the follow-up of the trajectory (treatment of short length ISO blocks abrupt changes of direction), vibratory behavior of the system Part/Tool/Machine, etc [Dugas et al. 2002]. In majority, these difficulties are mainly (but not exclusively) related to the use of the linear interpolation format [Lartigue et al. 2004]. One noticeable improvement is achieved by the use of polynomial formats whether they are performed off line [Lavernhe et al. 2004] or in real-time [Altintas et al. 2003] [Cheng et al. 2002].

Finally, the geometry of the machined part results from the tool movement calculated by the CAM system, altered by the behavior of the NC unit and deteriorated by dynamic phenomena during machining.

Therefore, in order to ensure the best possible performances in terms of quality and productivity, it is necessary to integrate a maximum of constraints during the step of tool path computation. The most complicated case is the case of 5-axis high speed milling. If 5-axis machining offers advantages compared to 3-axis machining (larger tool accessibility, better surface quality, greater rate of material removal), algorithms of tool path computation are more complicated.

Taking into account the two additional degrees of freedom, tool path computation methods lead to optimization problems of the tool axis orientation in order to satisfy a certain number of constraints. In majority, authors consider geometrical constraints in order to avoid interferences or collisions [Rao et al. 2000] [Jung et al. 2003]. More recent work attempts to integrate kinematical constraints related to the machine tool: smoothed orientations [Ho et al. 2003], limitation of rotary axis displacements [Munlin et al. 2004].

Let us notice that one of the main problems in 5-axis machining is that the tool trajectory, which is calculated in the part reference frame, is carried out in the machine tool reference frame through joint variables. It is thus necessary to take into account constraints such as the machine tool architecture, its kinematics, its axis ranges and the dimensions of its components.

This paper deals with kinematical performances of the machine tool for the optimization of the 5-axis tool path computation step. The study proposed relies on a machining model that allows integrating geometrical, kinematical and dynamical constraints as soon as possible in the step of tool path computation. The machining model is a surface model geometrically rich. Then, the surface model guarantees the use of a coherent geometrical model during the free-form machining process. It also facilitates the integration of constraints in order to optimize the tool path computation and ensures at least a second order continuity [Duc 1999].

Section 2 is dedicated to the presentation of the surface model for tool path computation. Section 3 emphasizes the importance of the inverse kinematical transformation in 5-axis machining. Finally, kinematical limits are integrated in tool path computation through an example in section 4.

## 2. A surface model for tool path computation

In order to minimize geometrical approximations appearing all along the free-form machining process, we proposed in a previous work a surface tool path model. The surface model ensures functional requirements and design intents while integrating machining constraints. Our first work led to the definition of the geometrical machining model as a parameterized surface called the *Machining Surface*. With this model, trajectory of a particular point of the tool is expressed as a surface or as a set of surfaces [Tournier et al. 2000].

In 5-axis machining, the tool axis orientation is specified by the user through two angles ($\theta_t$, $t$) and ($\theta_n$, $n$) in the local coordinate frame ($C_C$, $f$, $n$, $t$), where $f$ is the tool feed vector, $n$ the normal vector to the surface and $t$ the tangent vector to the surface defined by $t = f \wedge n$ (figure 1). For each cutter location, a point locating the tool and a vector to orient the tool axis must be specified. The *Machining Surface* thus consists of two surfaces $S_1$ and $S_2$ (figure 1), with $S_1$ the guiding surface and $S_2$ the orientation surface. The guiding surface $S_1$ is the locus of the point $K$ with $K = C_C + r.n$. It is thus the offset surface of the nominal surface by a distance equal to the corner radius $r$ of the tool. Therefore, the guiding surface does not depend on the machining strategy. The orientation surface $S_2$ is the locus of the point $C_L$ with $C_L = C_C + r.n + R.v$. It is thus this surface which gives the orientation of the tool axis according to the considered machining strategy.

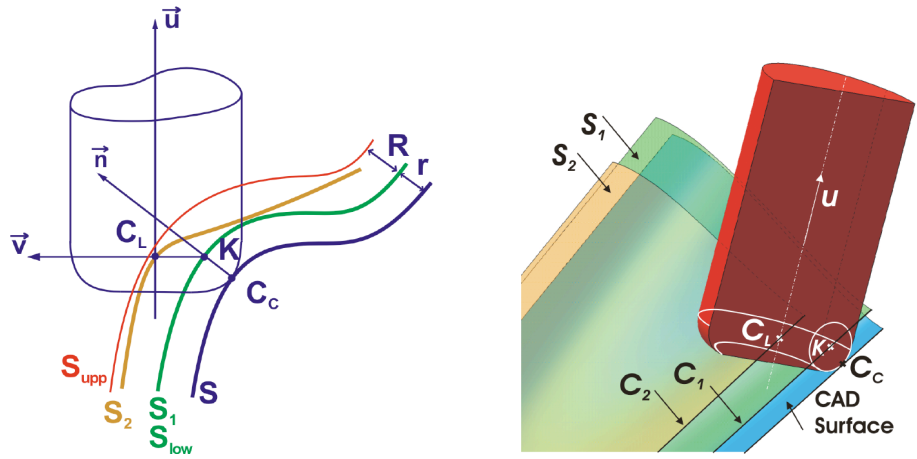

*Figure 1: the machining surface in 5-axis milling*

At this stage, the CAM model is described as a continuous surface which contains more information than the classical model made up of a set of ordered points.

The use of the machining surface ensures the continuity of the tool paths but also uncouples functional requirements and dynamical requirements. Indeed, the orientation of the tool axis only depends on the definition of the orientation surface. Therefore, the surface model allows integrating constraints, as constraints on the orientation surface, earliest in the tool path computation step. Constraints can be of different natures: machine tool geometrical constraints (axis range, part holder orientation), machine tool kinematical constraints (speed and acceleration on the 5 axes, singularities) and interference or collision constraints. For example, avoiding gouging can be performed by a deformation of the orientation surface which modifies the tool axis orientation consequently [Tournier *et al* 2000]. In other words, it exists a restriction of the 3D space where we can define the orientation surface to guarantee tool path from gouging.

This concept can be enlarged to other constraints linked to the tool axis orientation such as the instantaneous machining strip width [Lee *et al* 1997], [Tournier *et al* 2005], the chip thickness and the tool load [Hascoet *et al* 2000] or the kinematical behavior of the solicited machine tool axes. In this paper we will focus on this last point. Our objective is to define the shape and the position of the orientation surface in order to minimize the kinematical saturation of the machine tool axes in order to reach the programmed federate.

**3. Inverse kinematical transformation**

As our purpose is to anticipate the machine tool behavior during machining, it is necessary to control all the elements of the digital process from the CAD/CAM activities to the motor drives control by the NC unit. In particular, the post-processor plays a major role in 5-axis machining. It is important here to recall that in 5-axis machining, motions of the machine tool axes are different from the tool trajectory calculated in the part reference frame by the CAM software. Indeed, the CAD/CAM software calculates generic NC files (ISO 3592 CL files for instance), which are not directly understandable by the NC unit; NC unit only manages programs described in ISO 6983 format. A post-processor is then necessary to interpret programs. Furthermore, the tool path which is transmitted to the NC unit can be calculated in the part reference frame or in the machine tool joint space.

When calculated in the part reference frame, programs are independent of the machine tool structure used to machine the part. However, the NC unit has to perform in real-time the inverse kinematical transformation (IKT). When programs are calculated in the joint space, a post-processor dedicated to the considered machine tool is necessary to calculate the IKT in addition to language translation.

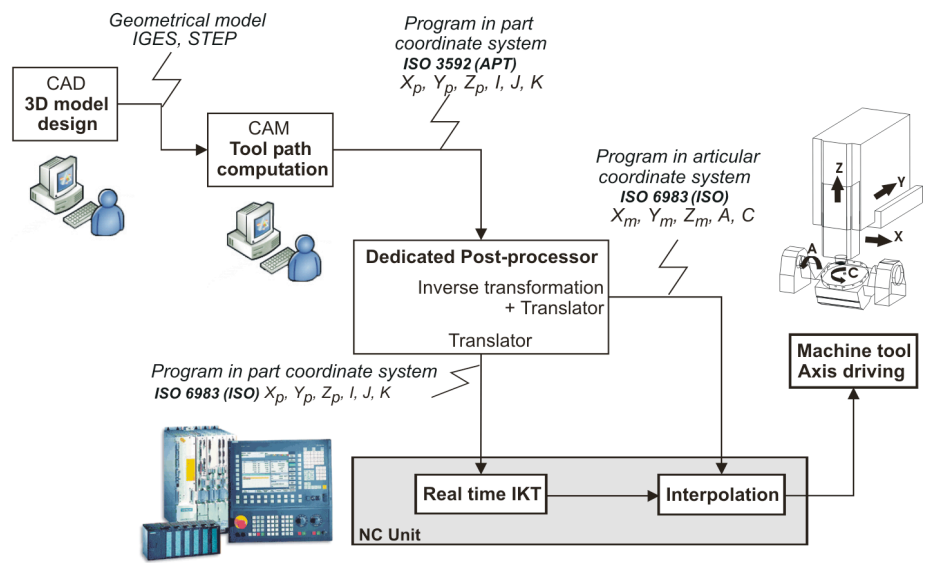

*Figure 2: post-processor in the digital factory*

The real time IKT performed by current NC units allows obtaining very satisfactory results [Duc et al. 2002]. Unfortunately, algorithms used are not detailed and accessible to machining users. In order to control the whole process, we have developed our own post-processor ensuring the IKT dedicated to the machine we use, a 5-axis vertical High Speed Milling center Mikron UCP710 with rotary axes A and C.

The development of IKT post-processors is confronted to two difficulties. The first one is linked to the multiplicity of solutions obtained after inversion of the coordinates. Indeed, one configuration in the part reference frame may correspond to several configurations in the machine tool joint space, even to an infinite number. This case may arise for singular points, and it is thus necessary to develop strategies to find out the solution less disturbing for the machining process [Affouard et al. 2004].

The second difficulty concerns the management of the tool trajectory in the part reference frame between two points of interpolation. Indeed, after reading the joint coordinate values at the ends of an elementary linear segment (more particularly angles A and C), the NC unit carries out a linear interpolation in the joint space. The resulting tool path in the part reference frame is thus unspecified. Furthermore, feedrate is not any more controlled. To overcome this problem, a solution is the use of the inverse time programming method: the time required to go from a point to another in linear interpolation is imposed to the NC unit. To proceed, it is necessary to determine the distance covered by the tool during its motion relative to the part [Poulachon 2004].

As this stage, we have a geometrical model giving the positions of the machine tool axes and a differential model giving axis velocities describing the movements of the tool relatively to the part.

## 4. Kinematical limits and application

The final objective of HSM 5-axis machining is the surface machining at high speed level while respecting the required geometrical quality. More generally, the geometrical quality is linked to geometrical specifications to be respected as well as the absence of marks on the part. The study of the cutting conditions provides the best feedrate for the couple tool/material considered.

For 5-axis machine tools, the rotary axes are those less powerful. If during machining these axes reach the maximum velocity admissible (saturation case), this involves the tangential programmed feedrate not to be reached. As a result, productivity is decreased and marks due to tool deflection may appear. Therefore, it is essential to generate optimal 5-axis tool path from a kinematical point of view.

For this purpose, we propose a process to determine optimal tool axis orientations (figure 3). This allows us to determine the best tool axis orientations without tests on the machine tool which will immobilize the means of production. We can notice that the choice of the tool axis orientation cannot be done without taking into account collisions, the part setup and the machining strategy.

The links between all these parameters makes the problem relatively complex to solve. The objective of the paper is not to completely solve the problem but to analyze links between the various constraints involved in 5-axis machining. In this direction, we make some simplifying assumptions.

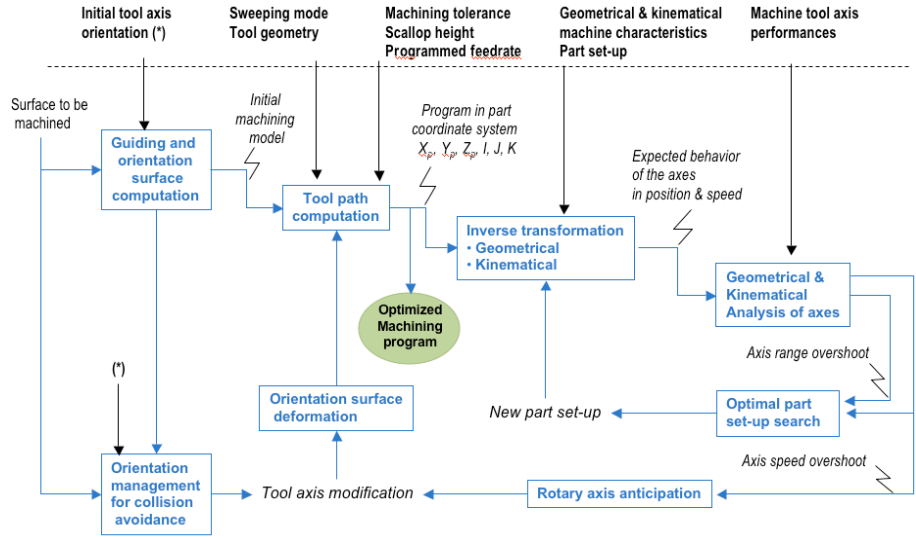

*Figure 3: tool axis orientation computation algorithm*

First, the part setup is imposed in the machine tool workspace. The position is centered on the machine table and the orientation is parallel to the machine axes. The machine tool is a Mikron UCP710 milling center with RRTTT kinematical structure. The surface to be machined is a hyperbolic paraboloid surface. The tool is a filleted end mill cutter tool. Tool path is calculated with a tool center guiding strategy, according to parallel planes oriented at 45° relatively to *X* and *Y* axes. The tool axis is oriented with a 1° tilt angle and a null yaw angle. The distance between guiding planes is fixed. The calculated tool path is presented in figure 4.

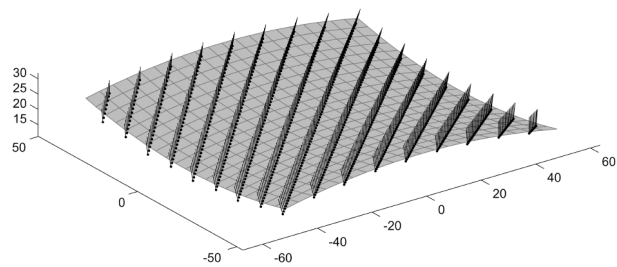

*Figure 4: test part and tool paths*

A preliminary study has validated that the considered tilt angle allows machining the part without collisions. The programmed feedrate is 5 m/min.

Supposing that the tool moves at the programmed feedrate, velocities of rotary axes are estimated using the inverse differential model, on three successive points of the tool path. The maximum velocity is 15 rpm for the *A*-axis and 20 rpm for the C-axis. We can see in figure 5 that *A*-axis solicitations remain within the axis admissible range whether not for the *C*-axis. Respecting the programmed feedrate during the trajectory follow-up leads to velocity saturation on the C rotary axis.

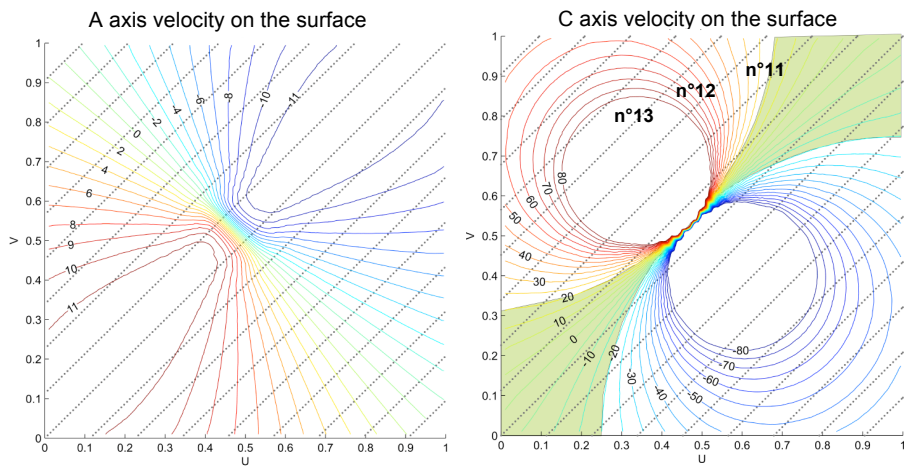

*Figure5: rotary axes velocity analysis*

We thus propose to modify the tool axis orientations in the area where the *C* axis saturates in order to reach the programmed feedrate. For this purpose, we analyzed the evolution of rotary axis velocities for various tilt angles. For some areas, a 5° tilt angle will generate less saturation, while preserving a 1° tilt angle for zones for which there is no solution. Consequently, we introduced local modifications for tool paths n°11, n°12 and n°13. This is carried out by deforming the orientation surface (Figure 6) imposing for some points of the path n°12 a 5° tilt angle. Thanks to the continuity of the orientation surface, a whole area is deformed leading to modify any position of the tool for the considered path.

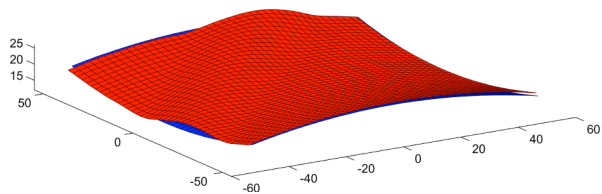

*Figure 6: deformation of the orientation surface*

In the transversal direction, the deformation involves a modification of the tilt angle on the preceding and following paths. We thus observe that velocity of the *A*-axis remains within the axis limits, and we can as far as possible limit saturations of the *C*-axis (figure 7).

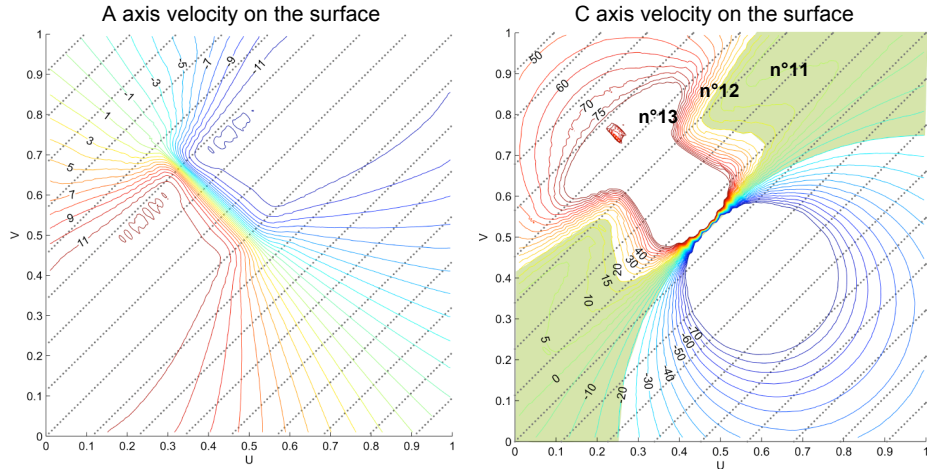

*Figure 7: rotary axis velocities after modification*

Modifications of the tilt angle have consequences on the machining strip width along the path. Indeed, when using a filleted end mill cutter tool which is oriented relative to the surface according to a tilt angle and a yaw angle, the effective profile of the tool in a perpendicular plane to the feed direction is an ellipse, the large radius of which, $R_{eq}$, is equal to:

$$Req = \frac{r(R + r\sin\theta t)}{r\sin\theta t \cos\theta n^2 + (R + r\sin\theta t)\sin\theta n^2}$$

When the yaw angle is null, this yields to:

$$Req = \frac{(R + r\sin\theta t)}{\sin\theta t}$$

We notice that for a null tilt angle, the equivalent radius $R_{eq}$ is infinite, which is not surprising for the tool consequently works with its flat bottom. On the other hand, collisions between the tool and the surface in the concave area are unavoidable. The numerical application shows that with a 5° tilt angle, the equivalent radius $R_{eq}$ is five times smaller than with a 1° tilt angle (104 mm instead of 516 mm). If the constant distance between parallel planes is preserved, the scallop height will be larger in the area for which we increased the tilt angle. Furthermore, in order to maintain geometrical variations within the limits fixed by tolerances, the tool path tightening in the deformed area can be automatically carried out according to the deformation of the orientation surface.

## 5. Conclusions and future works

The study presented in the paper allows predicting and analyzing the influence of the two additional degrees of freedom, the tilt angle and the yaw angle, used in 5-axis end milling. Starting from an initial solution for the tool axis orientation, we can predict the impact of modifications of the tool axis orientation on the kinematical behavior of the machine tool axes by means of a dedicated post-processor specifically developed for the used machine tool.

We have thus noticed that constraints linked to the tool axis orientation are antagonistic and should be subject to a global optimization. Indeed, a minimum tilt angle of the tool relative to the surface normal maximizes the machined strip width, while decreasing machining time.

On the opposite, such a tool axis orientation is the main source of collisions between the tool and the part. Lastly, modifications of the tool axis orientation in order to increase federate, i.e. to decrease machining time, will undoubtedly change the width of cut. As a result the number of tool paths will be increased and so machining time.

Current works are in progress considering the resolution of the global approach.

For tool path calculation activity, the choice of the machining strategy and of the tool axis orientation should not rise from an analysis a posteriori but must be the result of a synthesis of the constraints leading to an optimal solution. It is thus necessary to work on the expression of the kinematical constraints as geometrical characteristics in order to directly specify admissible tool axis orientations (that means the orientation surface).

Finally, we are now developing a simulator to predict the effective feedrate according to encountered geometrical difficulties, such as tangency and curvature discontinuities of tool paths, and taking into account the characteristics of the couple machine tool / NC unit used. Coupled with the post-processor, this simulator will allow a better understanding of the required positions and velocities on each machine tool axis.